\documentclass[aps,pre,reprint,nofootinbib,floatfix]{revtex4-1}
\usepackage{graphicx,amssymb,amsfonts,amsmath,color}

\newcommand{\hairpin}{
\begin{figure}[t]
\includegraphics[width=\columnwidth]{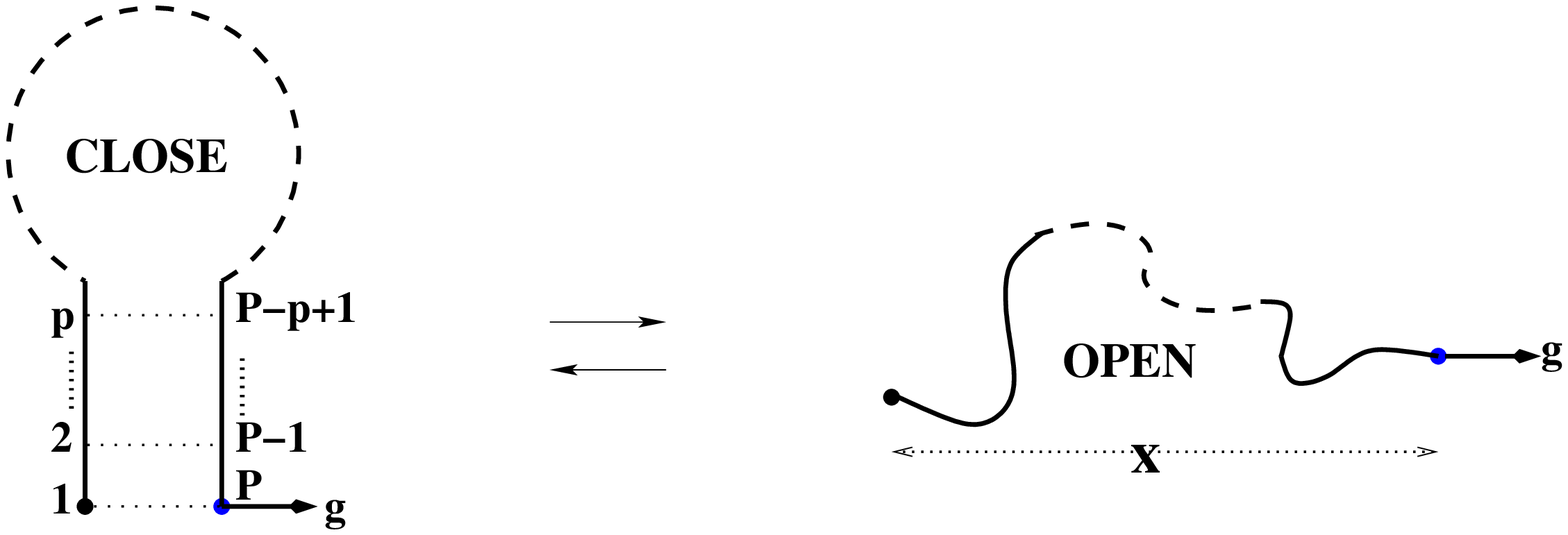}
\caption{ DNA hairpin in a zipped (Z) and an unzipped (U) state.  The
  stem (solid lines) consists of complementary nucleotides, whereas
  the loop (dashed line) is made up of non-complementary nucleotides.
  Only native interaction (dotted lines) for base pairing is allowed.
  For this paper, the stem is of length $p=10,$ and the loop has 12
  monomers. The total length is $P=32$.  }
\label{fig-1}
\end{figure}
}

\newcommand{\force}{
\begin{figure}[t]
\includegraphics[width=2.5in,clip]{PRE_fig2.eps}\hskip 0.75cm
\caption{ A typical periodic force used in the simulation.  A cycle
  refers to one period of variation of the applied force $g$ from $0$
  to $g_m$ (called amplitude) and back to $0$ at a fixed time period
  $\tau$.  This example has amplitude $g_m=0.1$ and two different
  frequencies $\nu=1/\tau=0.38$ and 0.19 for n=5 and 10 respectively.}
\label{fig-1-sup}
\end{figure}

}

\newcommand{\hyst}{
\begin{figure*}[t]
\hspace{-0.5cm}\includegraphics[width=\textwidth]{PRE_fig3.eps}
\caption{ (Color online) DNA hysteresis for different $g_m$ and $\nu$
  (increases from left to right) as indicated.  Each plot contains the
  loops for 10 different initial conformations.  These are at
  $T=0.10$, for which $g_c \sim
  0.20$.} 
\label{fig-2}
\end{figure*}
}

\newcommand{\pq}{%
\begin{figure*}[htbp]
\hspace{-0.1in}\includegraphics[width=\textwidth]{PRE_fig4.eps}
\caption{The time sequence of $Q$ and $P(Q)$ vs $Q$ plots.  Three
  pairs of vertical panels for three $\nu$'s.  Each panel has ({\it
    i}) the time sequence on the left (L), and ({\it ii}) $P(Q)$-plot
  on the right (R) for different $g_m$'s.  }
\label{fig3}
\end{figure*}
}

\newcommand{\pqvsq}{
\begin{figure*}[t]
  \includegraphics[width=\textwidth]{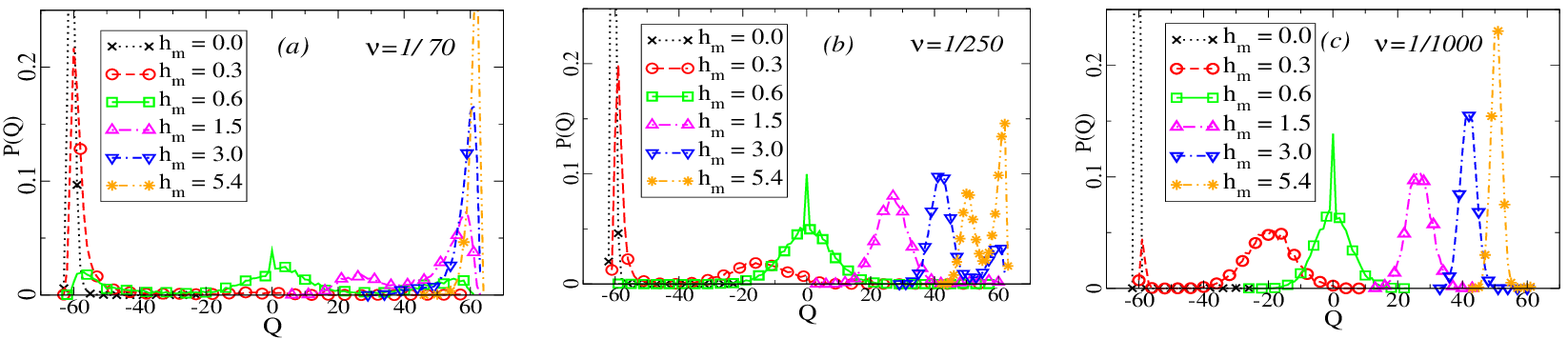}
  \caption{(Color online) $P(Q)$ vs $Q$ for the Ising model for three different
    $\nu$'s, and the same set of $h_m$ with $h_l=-0.6$. For a given
    $\nu$, the distribution $P(Q)$ shifts as $h_m$ increases. It goes
    from Z to U through D.  The sequence of phases are (a) Z
    $\rightarrow$ Z+D $\rightarrow$ Z+D+U $\rightarrow$ D+U
    $\rightarrow$ U.  (b)Z $\rightarrow$ Z+D $\rightarrow$ D
    $\rightarrow$ D+U.  (c) Phases: Z $\rightarrow$ Z+D $\rightarrow$
    D.}
\label{fig:pqvsq}
\end{figure*}
}

\newcommand{\bound}{
\begin{figure}[b] 
\includegraphics[width=\columnwidth]{PRE_fig5.eps}
\caption{(Color online) Dynamical phase diagram of a periodically
  driven DNA hairpin in the $g_m$-$\nu$ plane. The lines are
  boundaries for various phases U, D and Z. The points are from the
  simulation and the lines are guide to eye.  }
\label{fig:bound}
\end{figure}
}

\begin{document}
\title{Dynamical phase transition of a periodically driven DNA}
\author{Garima Mishra$^{1}$}
\email{garimabhu@gmail.com}
\author{Poulomi Sadhukhan$^{2}$}
\email{poulomi@iopb.res.in}
\author{Somendra M Bhattacharjee$^{2}$}
\email{somen@iopb.res.in}
\author{Sanjay Kumar$^{1}$}
\email{yashankit@yahoo.com}

\affiliation{$^1$Department of Physics, Banaras Hindu University,
     Varanasi 221 005, India. \\
$^{2}$Institute of  Physics, Bhubaneswar,  751 005, India.}

\begin{abstract}
  Replication and transcription are two important processes in living
  systems.  To execute such processes, various proteins work far away
  from equilibrium in a staggered way. Motivated by this, aspects of
  hysteresis during unzipping of DNA under a periodic drive are
  studied.  A steady state phase diagram of a driven DNA is proposed
  which is experimentally verifiable. As a two state system, we also
  compare the results of DNA with that of an Ising magnet under an
  asymmetrical variation of magnetic field.
\end{abstract}

\maketitle

\section{Introduction}

The semi-conservative replication of DNA requires a complete
separation of the two strands, which occurs concomitantly with the
production of new strands.  Even transcription needs a partial
separation of the two strands of DNA to read the genetic code of the
base sequence \cite{watson}. The traditional view of a temperature or
pH induced melting is getting superseded by a more mechanical
procedure of force or torque induced unzipping \cite{bhat99, helexpt,
  Bockelmann, heltheo}.  This paradigm shift is because of the lack of
extreme temperatures or the environment required in general for the
melting (the melting temperature is around 70-100C) and the recent
establishment of the unzipping as a genuine phase
transition\cite{bhat99}.  A double-stranded (ds) DNA remains in the
bound doublet state below its melting temperature even when pulled by
a force at one end until and unless the force exceeds a critical force
$g=g_c(T)$, above which it opens into two single stranded (ss) DNA.
Many aspects of this transition have been studied since then both
theoretically\cite{sebas,nelson,trov,kumar_prl,MukamelE,km,ps11,kap2,gi,kapri}
and experimentally\cite{Bockelmann,prentiss,hatch}, mostly in
equilibrium, though puzzles remain\cite{kumarphys,bundschuh}.

{\it In vivo}, a DNA is opened by helicases, which are motors that move
along the DNA\cite{watson}.  Both the motion and the opening processes
require constant supply of energy.  A few examples are DNA-B, a ring
like hexameric helicase that pushes through the DNA like a
wedge\cite{dnab}, PcrA that goes through cycles of pulling the ds part
of the DNA and then moving on the ss part\cite{pcra}, viral RNA
helicase NPH-II that hops cyclically from the ds to the ss part of DNA
and back\cite{nphii}.  Such cycles of action and rest, with the periodic
ATP consumption, indicates an exertion of a periodic force on the DNA.

It suffices to describe the equilibrium unzipping transition by the
two thermodynamic conjugate variables, force $g$ and separation $x$ of
the pulled base pair (Fig.1). In thermal equilibrium, a quasi-static
change of the force from zero to a maximum $g_m$ and then back to
zero, keeping other intensive quantities fixed, would result in
retracing the thermodynamic path, ending at the initial state; history
plays no role because of ergodicity.  In contrast, under a periodic
force, the mismatch between the relaxation time and the external
timescale for change of force would restrict the DNA to explore a
smaller region of the phase space, creating a difference in the
response to an increasing or a decreasing force.  Near a phase
transition, where the typical time scales of dynamics become large,
the difference between the forward and the backward branches becomes
prominent.  This is hysteresis of DNA unzipping\cite{mishra}.  More
recently Kapri\cite{kapriWT} showed how the work theorem can be used
{\it via} a multi-histogram method to extract the equilibrium isotherm
from the hysteresis curves.  The nature of hysteresis and its dependence
on the applied field and frequency are well studied in the context of
magnetic and structural systems\cite{bertotti,bkcrmp}, though it is
the time-averaged loop that has received the attention.  With the
advent of single-molecule experiments on short DNA chains (oligomers),
it might be possible to probe the time-resolved loops, not just
averages.  Motivated by the biological relevance and the experimental
feasibility, we explore the behavior of DNA under a periodic force, to
be called a {\it periodically driven DNA}.  Our results show that
without changing the physiological conditions ({\it e.g.}  the
temperature or pH of the solvent), a DNA chain may be brought from the
unzipped state to the zipped state and {\it vice versa} by varying the
frequency ($\nu$) alone.  By using a probabilistic description of the
time variation of the DNA response, we propose a force-frequency
$(g$-$\nu)$ phase diagram for the driven DNA.

A similar dynamic symmetry breaking transition is known to occur in
magnets where the nature of the hysteresis loop changes\cite{bkcrmp}.
Because of the two stable phases, the unzipping transition has often been
described by a two state Landau type free energy functional
\cite{heltheo,ps11,gupta}.  As a two phase system, we make use of the
magnetic Ising model to corroborate the behavior of DNA, both
undergoing a first order transition and showing hysteresis.  We
establish that the observed features and the phase diagram are robust
and generic, and not tied to any particular DNA model.

It was recently shown that the DNA hairpin (Fig. 1) of 32 beads 
(stem 10 base pairs and loop of 12 bases) captures the   essential 
properties of a long DNA chain with implicit bubbles \cite{mishra}.  The
force-temperature diagram of the DNA hairpin is found to be  qualitatively 
similar to the DNA unzipping experiment \cite{prentiss} in the entire range 
of $f$   and $T$, whereas the phase diagram for a dsDNA of 16 base pairs 
differ significantly with  the experiment. Since, under the periodic force, 
the chain may unzip and rezip, the  bubble dynamics and the wandering 
interface (Y-fork junction) meeting   transient bubbles (extra source of 
entropy), would play important roles in hysteresis.  Needless to
say that such bubbles are ubiquitous in a long chain. A DNA
hairpin is the simplest model, where the premade loop allows us to
study the dynamics of such implicit bubbles.

The outline of the paper is as follows.  In Sec.
\ref{sec:dna-hairpin-under}, we introduce the model used for the
hairpin and the method of study. The results are analyzed in terms of
average hysteresis loops and time resolved loops.  The dynamic phase
diagram can be found in this section.  In Sec
\ref{sec:dynam-phase-trans}, the hysteresis loops of an Ising
ferromagnet under an asymmetric field modulation is given.  The
similarities with the DNA hairpin problem are discussed here.  The
paper ends with a short summary and conclusion in Sec.
\ref{sec:summary}.

\hairpin

\section{DNA hairpin under periodic force}\label{sec:dna-hairpin-under}

\subsection{Model and method}

DNA hairpins (see Fig.\ref{fig-1}) occur naturally {\it in vivo} and
are used often in experiments on DNA with bubbles \cite{bonnet}.  The
non-paired bases of the loop is a source of entropy which in turn
controls the dynamics of hysteresis. In this paper, we perform
Langevin dynamics (LD) simulations of a DNA hairpin\cite{mishra},
to study the separation $x$ of the terminal base pairs pulled by a
periodic force $g(t)$ of time period $\tau (=1/\nu)$ (Fig.
\ref{fig-1-sup}). There have been efforts \cite{kap2,gi} to
understand transcription like processes by applying a constant force in the
middle of a DNA chain.  A very rich phase diagram has been obtained
though no such experiment has been reported so far to confirm the
phase diagram. It appears that applying a force in the middle of the chain is
experimentally difficult. Moreover, the biological situation requires
the force following a moving replication fork, which is also not
amenable to single molecule force experiments. In view of these, we
use a simpler geometry where the force is applied at one end of the hairpin
keeping the other end fixed, which could be used in single molecule experiments. 
The typical time scale of fluctuations
involved in unzipping varies from $ns$ to $\mu s$, and therefore we avoid
the details of an all atom simulation by adopting a minimal
off-lattice coarse grained model of a DNA hairpin, where each bead
represents nucleotide as a basic unit, which comprises a base, a sugar
and a phosphate group. We have performed our simulation in reduced unit as
discussed below.

\subsubsection{Model}
The configurational energy of the system under consideration is\cite{mishra}
\begin{eqnarray}
 E & = & \sum_{i=1}^{P-1}
 k(d_{i,i+1}-d_0)^2+\sum_{i=1}^{P-2}\sum_{j>i+1}^{P}4(\frac{C}{d_{i,j}^{12}}- 
\frac{A_{ij}}{d_{i,j}^6}).
\label{eq1}
\end{eqnarray}
Here, $P=32$ is the number of beads. The harmonic term with a spring
constant $k$ (=100) couples the adjacent beads along the chain.  The
distance between the adjacent beads $d_{i,j}$'s defined as $|{\bf
  r}_i-{\bf r}_j|$, where ${\bf r}_i$ and ${\bf r}_j$ denote the
positions of beads $i$ and $j$, respectively. We assign $C=1$ of the
Lennard-Jones potential. The base pairing interaction $A_{ij}=1$ is
restricted to native contacts only, which is similar to the Go model.
All pairs of nucleotides that do not form native contacts in the stem
and in the loop interact only through short range repulsion (excluded
volume).  The parameter $d_0 (=1.12)$ corresponds to the equilibrium
distance in the harmonic potential, which is close to the equilibrium
position of the average L-J potential. We obtained the dynamics of the
system by using the following Langevin equation \cite{Allen,Smith}
\begin{equation}
\label{eq:1}
m\frac{d^2{\bf r}}{dt^2} = -{\zeta}\frac{d{\bf r}}{dt}+{\bf F_c}+{\bf \Gamma},  
\end{equation}
where $m$ is the mass of a bead which is set equal to one here, and
the friction coefficient used in simulation is $\zeta =0.4$. ${\bf
  F_c}$ is defined as the gradient of the energy, $-\nabla E$, and
${\bf \Gamma}$ is a random force, a white noise with zero mean and
correlation
\begin{equation}
  \label{eq:4}
 <{\Gamma_i(t)\Gamma_j(t')}> = 2\zeta k_B T\delta_{ij}(t-t'),
\end{equation}
where $k_B$ is the Boltzmann constant which is set to one in present
simulation.  The choice of Eq. \ref{eq:4} keeps the temperature of the
system constant through out the simulation for a given force.  In
order to study the behaviour of a DNA hairpin under a periodic force,
we add an additional energy $-g(t).x(t)$ to the total energy of the
system Eq.\ref{eq1}. We have performed the simulation at $T=0.10$, for
which $g_c \sim 0.20$ \cite{text2}.  We may convert the dimensionless
units to real units by using the relations:\cite{Allen},
\begin{equation}
  \label{eq:5}
 t^*= \left(\frac{m {\sigma}^2}{\epsilon}\right)^{1/2} t,\  
  r^*=\sigma r,
\end{equation}
where $t^*$, $r^*$ and $\epsilon$ are time, distance and
characteristic hydrogen bond energy in real units.  $\sigma$ is the
distance at which inter-particle potential goes to zero. For example,
if we set effective base pairing energy $\epsilon \sim 0.1$ eV, the
average mass of each bead $5 \times 10^{-22}$ g and $\sigma=5.17$ \AA,
we get a time unit $\tau \approx 3$ ps.  It is known that $\zeta$
affects the kinetics only and thermodynamics remain unchanged
\cite{heyon,Li_bpj}.  The friction coefficient used in our simulations
$\zeta = 0.4 m\tau^{-1}= 6 \times 10^{-11} {\rm g s}^{-1}$.  The order
of force thus would be $\epsilon/$\AA $\sim 160$ pN. For temperature
conversion, a little more care is needed because the coarse grained
model does not take into account many other energies\cite{Kibbe}.  We
compare the phase diagrams for DNA in the thermodynamic limit using a
two state model based on the modified Freely Jointed Chain (mFJC)
model of polymer with the experiment \cite{prentiss}.  The melting
temperature at $363$K and the flat portion (the stretching of covalent
bonds) at $293$K in experiments\cite{prentiss} correspond to $0.23$
and $0.05$ in reduced units, which are also consistent with the
simulation \cite{mishra}.  If we use this information to scale the
temperature in a linear fashion, we have a relation, $T^*= 363+
\frac{363-293}{0.23-0.05} (T-0.23)$, where $T^*$ is the real
temperature.  This mapping is valid only in the range of overlap of
mFJC and experiment.  This relation gives the reduced temperature used
here $T=0.1$ as equivalent to a real temperature around $320$K, i.e.,
around 45C. It may be noted that hysteresis has been observed in this
range of temperature \cite{hatch}.

\force

\subsubsection{Integration}
The equation of motion is integrated using a sixth order predictor
corrector algorithm with a time step $\Delta t$=0.025 \cite{Smith}.
Ref. \cite{mishra} showed that the present coarse-grained model
exhibits the observed features of the experimental $g$-$T$ diagram
\cite{prentiss} of a long DNA with implicit bubbles as well as
hysteresis at low temperatures in the $g$-$x$ plane as seen
experimentally \cite{hatch}.  During the simulation, $g$ is changed in
steps of $\Delta g (= 0.01)$ from $0$ to a maximum $g_m$ and then to
$0$ (Fig.  \ref{fig-1-sup}).  This one period is to be referred to as
{\it a cycle} and $g_m$ as the amplitude.  The time period 
$$\tau=2 n \Delta t (1+ \frac{g_m}{\Delta g}),$$ 
is controlled by $n$, the number of LD steps executed after every
increment (or decrement) of force.  We choose $n$ such that
$n \Delta t$ is much below the equilibration time.

By changing $g_m$ or $\nu$, we find it is possible to induce a
dynamical transition between a state of time averaged zipped (Z) or
unzipped (U) to a dynamical state (D) oscillating between Z and U.

\subsection{Numerical analysis}

\subsubsection{Hysteresis loops}

In Fig.\ref{fig-2}, we plot the average value of $x(g)$ over
${\cal{C}}~(=1000)$ cycles vs $g$, for different values of $g_m$ and
$\nu$.  These loops for different initial conformations remain almost
the same (except (23)) so that the large menagerie of shapes of
hysteresis loops observed are typical, not exceptional or accidental.
At a high frequency, the DNA remains in Z or in U depending on whether
$g_m <2g_c$ or not, (Fig.\ref{fig-2}(13)\&(43)), irrespective
of the initial conformation.  For a relatively smaller $\nu$, the
sequence of hysteresis loops (Fig.\ref{fig-2} (11,21 \& 31)) is of a
different nature.  These clearly reflect that the DNA starts from the
Z ($x = 0$ at the start of the cycle), goes to the U ($x = 30$) and
then back to the Z \cite{text3}.  Some of these plots show the phase
lag between the force and the extension, e.g., even when the applied
force decreases from $g_m$ to 0 (Fig.  \ref{fig-2} (32 \& 33)), the
extension $x(g)$ increases.  If the system gets enough time to
approach equilibrium, the lag disappears (e.g., Fig.\ref{fig-2}(21 \& 31)).
A different behavior is seen in the case of intermediate forces
(Fig.\ref{fig-2} (21, 22 \& 23) ).  Despite the identification of the
states at the two extreme forces as Z and U, there is a significant
change in the $x$-values at $g=0$ and $g=g_m$, in Fig.\ref{fig-2}(23),
compared to the other loops shown in Fig.\ref{fig-2}.  Most striking
here is the wide sample to sample fluctuations in the loop.

\hyst

\subsubsection{Time-resolved loops}

The failure of the average response to provide a description of the
steady state dynamic behavior prompts us to analyze the distribution
of paths over the different cycles in terms of a dynamical order
parameter $Q$ defined as
\begin{equation}
Q=\frac{1}{\tau}\int_0^{\tau} x(t) \, dt.
\end{equation}
This $Q$ is analogous to the dynamical order parameter studied in the
context of magnetic systems\cite{bkcrmp}.  In Fig.\ref{fig3}, we plot
the values of $Q$ over different cycles for three different $\nu$.
The time sequence of $Q$ seems not to indicate any regular pattern,
and therefore, we assume that the allowed states occur randomly.  The
time sequence can then be interpreted in terms of a probability of
getting a particular value of $Q$.  We find that the steady state is
described by a stationary probability distribution ($P(Q)$) which are
also shown in Fig \ref{fig3}. 
The plots show three peaks which could be associated with phases U, D
and Z by comparing with Fig. \ref{fig-2}.  Based on the width of
$P(Q)$ we make operational criteria, $0<Q_Z<10$, $Q_U>22$, and
$10<Q_D<22$ for the identification of the phases.

\pq
 
\bound

From the peak locations of $P(Q)$, we map out the phase diagram of the
driven DNA in the $g$-$\nu$ plane (Fig~\ref{fig:bound}).  A line in
Fig.\ref{fig:bound} represents a boundary beyond which a particular
peak appears or disappears and resembles a first-order line. These are
now discussed sequentially.

{\it{(A)}} For a very small force, $P(Q)$ expectedly shows a peak in
the $Q_{{\rm Z}}$ window for any $\nu$ (Fig.~\ref{fig3} 1st row).
Beyond a certain $g_m$ there appears a second peak in the $Q_D$
window.  See Fig.~\ref{fig3} 12 to 22 or 13 to 23.  For each chosen
$\nu$, we determined the value of $g_m$ above which the peak of $Q_D$
appears, giving us the lower boundary for D.

{\it{(B)}} For a given $\nu$, with increasing $g_m$, the peak height
of $Q_{{\rm Z}}$ decreases as seen in Fig.~\ref{fig3} along any
vertical sequence, top to bottom.  One gets one or two other peaks but
Z disappears beyond a certain $g_m$, giving an upper boundary for Z.

{\it{(C)}} Before Z disappears, there could be peaks for D or U, or
both ( Fig. \ref{fig3}(31, 22)), implying coexistence, Z+U or Z+D+U,
for a range of $g_m$.  From the appearance of U, we determine the
lower boundary for U.

{\it{(D)}} Then, as Z disappears, the system may be either in the D
phase (Fig.~\ref{fig3}, 23R), or, in the mixed state of D+U
(Fig.~\ref{fig3}, 32R).  To be noted is the crossing of the boundaries
of U and Z, that produces the region of D phase.  Once the mixed phase
D+U appears, a further increase of $g_m$ would vanish the D peak and
only U peak survives.

An important point is the occurrence of different $Q$ values or regions
of phase coexistence.  In such a situation, the average response is
not a good measure of the state and there will be intrinsic difficulty
in reducing fluctuations in the hysteresis curve.

\pqvsq

\section{Dynamical phase transition in Ising ferromagnet}\label{sec:dynam-phase-trans}

To use the magnetic analogy, consider a ferromagnet like a
two dimensional Ising model below its critical point, subjected to a
periodic magnetic field $h=h_0 \cos (2\pi\nu t)$ in time ($t$).  As
$\nu$ increases, there is a dynamic transition where the average
magnetization in a cycle goes from zero to a nonzero
value\cite{bkcrmp}.  For small $\nu$ there is a hysteresis loop
connecting the two symmetrically opposite magnetized states, while for
large $\nu$, the lethargic system remains magnetized in one direction.
The analogy is between the conjugate pair $(h,m)$, where $m$ is the
magnetization, for a magnet and $(g,x)$ for DNA.  The magnetic
hysteresis loop in the $h$-$m$ plane is analogous to that in
the $g$-$x$ plane.  To simulate a DNA-like behavior in the Ising
system, an asymmetric modulation of the magnetic field has been taken
to make the average field over a cycle different from the critical
value.  A Monte Carlo dynamics is used to study the hysteresis in an
$8\times 8$ square lattice Ising model with periodic boundary
conditions under a periodic field between $h_l$ and
$h_m$.

\subsection{Model and Method}
We consider a two-dimensional Ising spin system ($\{s_i=\pm 1\}$) with
nearest neighbor interaction and under a magnetic field $h$,
\begin{equation}
  \label{eq:2}
  H = -J\sum_{\langle i,j\rangle} s_is_j-h\sum_{i}s_i, \qquad (J>0), 
\end{equation}
with $i$ labeling the sites of an $8\times 8$ square lattice with
periodic boundary conditions. The infinite lattice critical
temperature is $T_c\approx 2.269 J/k_B$ in zero field.  The
magnetization is defined as the thermal average
\begin{equation}
    \label{eq:3}
    m=N^{-1} \sum_i  \langle s_i\rangle,
\end{equation}
of $N (=64)$ spins.  For the above Hamiltonian, we choose $J=1$ and $k_B
T=2$ with $k_B=1$, so that $T<T_c$.

The Monte Carlo procedure used is as follows. We choose a spin,
calculate the change in energy $\Delta E$ of the system if the spin
gets flipped. Whether this spin would be flipped or not is determined
by using the standard Metropolis algorithm, which gives the
probability of acceptance of an attempted flip by
{\hbox{$P_{\rm{accept}}={\rm min}(1,e^{-\beta \Delta E}).$}} In this
way, we sequentially consider all the $N$ spins, one at a time, to
attempt a flip. The time taken to attempt $N$ spin flips constitutes
one MC sweep.

A complete cycle consists of $2{\cal N}$ steps, starting from $h_l$,
reaching $h_m$ and then back to $h_l$. Initially system is
equilibrated at $h_l=-0.6$ and $k_BT=2$. Then the periodic magnetic field
is switched on. At each step, (i) the magnetic field is increased by
$\delta h=(h_m-h_l)/{\cal N}$ and (ii) the system is relaxed towards
equilibrium by $n=5$ MC sweeps, which is much less than the
equilibration time. The magnetization $m$ is calculated at each of
such $2{\cal N}$ steps. The average of magnetizations calculated over
a cycle then gives the quantity $Q$. The above process is repeated
${\cal C}=10^4$ times, i.e., for $10^4$ cycles. The $Q$ values
obtained in this way gives the probability distribution $P(Q)$ for a
given $h_m$. We simulate the system for various frequencies
(controlling ${\cal N}$) and $h_m$, keeping $h_l$ fixed.

An asymmetric field in the Ising model enables us to distinguish two
differently ordered phases, the counterparts of the zipped and the
unzipped states, and in addition a {\it hysteretic state}, to be
called the {\it dynamic state} D.  For easy comparison, the negatively
magnetized state is named Z while the positively magnetized state is
U.  The states are distinguished by a dynamic parameter,
\begin{equation}
  Q=\frac{1}{\tau} \int_0^{\tau} m(t) \, dt,
\end{equation}
and the operational definition adopted is (i) $Q_{\rm Z}\equiv
\{-64\le Q\le -40\}$, (ii) $Q_{\rm U}\equiv \{55\le Q\le 64\}$, and
(iii) the rest is $Q_{\rm D}$. Cases (i) and (ii) occur when the paths
in the $m$-$h$ diagram remain on one side throughout the cycle, and
(iii) the dynamical state, D, occurs either if there are paths
connecting positive and negative values of magnetization or if the
paths remain more or less near zero of magnetization.  The division of
the three intervals or regions are independent of $h_{m}$.

\subsection{Numerical results}

The results for $P(Q)$ vs $Q$ shown in Fig. \ref{fig:pqvsq} are very
similar to Fig. \ref{fig3} and the interpretations toe the DNA line.
We point out a few salient features here.  For $h_m\le 0$, there is
only the $Z$ state (Fig. \ref{fig:pqvsq}abc).  With increasing $h_m$,
$P(Q)$ develops a multi-peaked structure which indicates coexistence.
Various sequences of states seen are noted in the figure caption.
These include D+U, Z+D and Z+D+U.  We find a clear dependence on $\nu$
of the occurrence of the various states, similar to the DNA case.
There are situations where U appears before Z vanishes.  In this
regime, we see the coexistence of three phases, Z+D+U (e.g., $h_m=0.6$
in Fig.\ref{fig:pqvsq}).  The peak height for D first grows from zero
and then depending on frequency, it may decrease, with a gradual
appearance of a U peak (Fig.  \ref{fig:pqvsq}a,b), or it may keep on
increasing up to a very high field seemingly merging with U.  (low
frequency case of Fig. \ref{fig:pqvsq}c).  The Ising case has a
special line at $h_m=-h_l$ where the $+\leftrightarrow -$ symmetry is
restored. On that line, there is the dynamic transition
(D$\leftrightarrow$ Z or U) at a particular frequency\cite{bkcrmp}.
The occurrence of the Z+D+U coexistence region found on this line is a
finite size effect and is a novelty of the mesoscopic system.  The
$h$-$\nu$ phase diagram is qualitatively similar to that of the DNA
hairpin, shown in Fig \ref{fig:bound}.

\section{Summary}\label{sec:summary}

This paper reports the possibility of a dynamical transition of a DNA
hairpin and a magnetic spin system both under a periodic drive.  When
a DNA is subjected to a periodic force, or, a magnetic system is
subjected to a periodic magnetic field, then the system driven away
from equilibrium shows hysteresis. Usually to get the hysteresis
curve, one averages over many cycles of the field or force. But we
show here that this averaging suppresses or loses the actual picture
of the states.  This can be observed if one looks at the individual
cycles. By quantifying the state of a cycle by $Q$, the average
separation for DNA or the average magnetization for the spins, we find
that the distribution of $Q$ is multi-peaked. Whenever a distribution
is multi-peaked, the average is not a good representation of the
state.  We identify the peaks of the distribution with the zipped (Z),
unzipped (U) and the dynamical (D) state for the DNA.  There is a
dynamical transition from one phase or state to another phase, and
also there are coexistence of phases in some range of strength of
drive and frequency.  The phase diagrams constructed from the data for
both the DNA hairpin and the magnetic system are of similar nature.
The role of hysteresis in biological processes remains an unexplored 
territory.  On a more quantitative front, our work calls for time 
resolved experiments on periodically driven DNA hairpins to explore such 
hitherto unknown dynamical phase diagrams.

GM and SK would like to thank CSIR and DST, India for financial
supports.


\pagebreak


\begin{thebibliography}{99}
\bibitem{text2}
The value $g_c \sim 0.2$ is from Fig. 4a of \cite{mishra}, marked by an
arrow there. We have chosen $T=0.1$  because hysteresis starts occurring 
as noted there.
\end{thebibliography}
\end{document}